\documentclass{article}
\usepackage{amsmath,graphicx,mlspconf}
\usepackage{amssymb}
\usepackage{algorithm2e}
\RestyleAlgo{ruled}
\usepackage{bbm}
\usepackage{multirow}
\usepackage{array}
\usepackage{tikz,pgfplots,pgfplotstable}
\pgfplotsset{compat=1.10}
\usepgfplotslibrary{groupplots,fillbetween}
\usetikzlibrary{intersections,calc,arrows,matrix,spy}
\usepackage{color}
\definecolor{darkred}{rgb}{0.6,0,0}
\definecolor{darkgreen}{rgb}{0,0.5,0}
\definecolor{darkblue}{rgb}{0,0,0.5}
\definecolor{SkyBlue}{rgb}{0.53, 0.81, 0.92}
\pgfplotsset{compat=1.5.1}

\usepackage{bm, url, hyperref}
\usepackage{cite}
\usetikzlibrary{shapes,arrows}
\usetikzlibrary{tikzmark}
\usepackage{color}
\DeclareMathOperator*{\argmin}{arg\,min}

\copyrightnotice{U.S.\ Government work not protected by U.S.\ copyright}

\copyrightnotice{979-8-3503-2411-2/23/\$31.00 {\copyright}2023 Crown}

\copyrightnotice{979-8-3503-2411-2/23/\$31.00 {\copyright}2023 European Union}

\copyrightnotice{979-8-3503-2411-2/23/\$31.00 {\copyright}2023 IEEE}

\toappear{2023 IEEE International Workshop on Machine Learning for Signal Processing, Sept.\ 17--20, 2023, Rome, Italy}

\def\M{{\mathbf M}}
\def\S{{\mathbf S}}
\def\L{{\mathbf L}}
\def\Q{{\mathbf Q}}
\def\U{{\mathbf U}}
\def\V{{\mathbf V}}
\def\R{{\mathbf R}}
\def\I{{\mathbf I}}

\def\RR{{\mathbb R}}

\def\T{{\mathcal T}}
\def\H{{\mathcal H}}
\def\P{{\mathcal P}}

\DeclareMathOperator{\rank}{rank}

\title{Deep Unrolling for Nonconvex Robust Principal Component Analysis}
\name{Elizabeth Z. C. Tan$^1$,  Caroline Chaux$^2$, Emmanuel Soubies$^3$, and Vincent Y.~F.~Tan$^{1,4}$\thanks{This work was supported by the National Research Foundation, Prime Minister’s Office, Singapore under its Campus for Research Excellence and Technological Enterprise (CREATE) Programme.}  }

\address{$^{1}$ Department of Mathematics, National University of Singapore, Singapore \\
     $^{2}$ CNRS, IPAL, Singapore $\qquad$
     $^{3}$ CNRS, IRIT, Université de Toulouse, Toulouse, France\\
     $^{4}$ Department of Electrical and Computer Engineering, National University of Singapore, Singapore}

\begin{document}

\maketitle

\begin{abstract}
We design  algorithms for  Robust Principal Component Analysis (RPCA) which consists in decomposing a matrix into the sum of a low rank matrix and a sparse matrix. We propose a deep unrolled algorithm based on an accelerated alternating projection algorithm which aims to solve RPCA in its nonconvex form. The proposed procedure combines  benefits of  deep neural networks and the interpretability of the original algorithm and it automatically learns   hyperparameters. We demonstrate the unrolled algorithm's effectiveness on synthetic datasets and also on a face modeling problem, where it leads to both better numerical  and visual performances.
\end{abstract}
\begin{keywords}
RPCA, Sparsity, low-rank, unrolled algorithm, hyperparameters.
\end{keywords}
%
\vspace{-.1in}
\section{Introduction}
\label{sec:intro}
\vspace{-.1in}
Robust Principal Component Analysis (RPCA) is the task of recovering a low rank matrix $\L^\star \in \RR^{d_1 \times d_2}$ and a sparse matrix $\S^\star \in \RR^{d_1 \times d_2}$ from their linear combination~\cite{candes2011robust} 
\begin{equation}\label{eq:model}
    \M^\star = \L^\star + \S^\star.
\end{equation}
Finding an exact solution to the RPCA problem is  challenging due to its combinatorial nature. Yet, RPCA has received considerable attention due to its   importance in many fields. These include applications from {\em latent semantic indexing}~\cite{deerwester1990indexing} to {\em image processing}~\cite{facerec:2018}, to {\em learning graphical models with latent variables}~\cite{chandrasekaran}, and to  {\em collaborative filtering}~\cite{koren2021advances}. \vspace{-0.2cm}

\paragraph*{The art of conventional RPCA:} Some authors \cite{rpcaconvex:2009}, \cite{rpcaconvexlista:2009} considered a convex relaxation of RPCA, where the low rank matrix is obtained throughout the minimization of the nuclear norm and the sparse matrix via an $\ell_1$-norm penalization. Such optimization problems can be solved by proximal gradient methods. However, such approaches are computationally expensive due to the proximal mapping of the nuclear norm, which involves a full singular value decomposition (SVD) of a $d_1 \times d_2$ matrix, amounting to at least $\mathcal{O}(d_1 d_2 \min(d_1, d_2))$ flops per iteration. 
In contrast, alternating algorithms have been proposed to solve the original nonconvex formulation of RPCA involving the $\ell_0$ pseudo-norm and the rank function (Section~\ref{sec:Alg_RPCA}). These include the alternating projections (AltProj) method \cite{altproj:2014}, its accelerated version (AccAltProj) \cite{accaltproj:2019}, and a block-based method based on the CUR decomposition~\cite{ircur:2021}. Although faster and more closely related to model~\eqref{eq:model} compared to methods based on convex relaxations, their performance heavily rely on good initializations. \vspace{-0.2cm}

\paragraph*{Learning-based strategies in RPCA:} Deep neural networks (DNNs)  have experienced a surge in popularity over the past decades, often attaining groundbreaking performance in various applications. In  signal processing, incorporating deep learning approaches has become prominent because of their ability to automatically learn salient information from  of real world data. However, DNNs are known to suffer from two   shortcomings. Firstly, their black-box nature (i.e., the lack of interpretability) hinders our understanding of why certain predictions are derived, which is crucial in detecting limitations. Secondly, they are susceptible to overfitting to the  training data since they often have a large number of  parameters compared to the  amount of available   training data.

To overcome these limitations, a technique known as \textit{deep unrolling} (also known as \textit{deep unfolding}) has been extensively explored \cite{Monga2021} and has emerged as a promising approach in various signal processing problems. While the model parameters are fixed in the classical algorithms, the unrolled network replaces them with learnable parameters that can be optimised through end-to-end training using backpropagation. Therefore, a trained unrolled network can be viewed as a parameter-optimised algorithm, sharing both the benefits of conventional DNNs and interpretability of the original algorithm. Furthermore, as classical algorithms often have significantly fewer parameters than DNNs, unrolled networks can potentially mitigate the overfitting problem when there is insufficient data or when the  training dataset is of low quality.

Existing unrolling strategies in the context of RPCA are currently limited to algorithms based on  convex relaxations. These include CORONA \cite{corona:2020}, refRPCA \cite{refrpca:2021}, and other similar works \cite{solomon:2019}, \cite{ultrasound:2020}, \cite{xray:2022}, \cite{multimodalrpca:2022}. However, they inherit from the  previously mentioned drawbacks of such convex relaxations. To the best of our knowledge, there does not exist unrolled versions of the alternating projections algorithm in RPCA, despite that being the state-of-the-art.
Indeed, such an unrolled algorithm be  beneficial in terms of having appealing computational properties and the closeness to model in~\eqref{eq:model} of nonconvex RPCA approaches, while mitigating existing shortcomings (sensitivity to the initialization and incognizance of hyperparameters).


\textbf{Contributions:} We propose an unrolled version of the Accelerated Alternating Projections algorithm \cite{accaltproj:2019}. The proposed procedure also incorporates  the Minimax Concave Penalty (MCP), an alternative to hard thresholding owning numerous interesting properties and more suitable than the $\ell_1$-norm relaxation \cite{lrpca:2021}.
The overall proposed procedure performs excellently on benchmark synthetic datasets and real-world (face) datasets, exceeding the performances on  the state-of-the-art (unrolled) approaches. 

\textbf{Outline:} The paper is organised as follows. Preliminaries on RPCA algorithms and deep unrolling are presented in Section \ref{sec:prelim}. The proposed method is then described in Section~\ref{sec:prop_algo} and numerical experiments are conducted in Section~\ref{sec:simu}. Finally, concluding remarks are presented in Section~\ref{sec:conclu}.

\vspace{-.12in}
\section{Prelimaries}
\label{sec:prelim}

\vspace{-.07in}
\subsection{Algorithms for RPCA}\label{sec:Alg_RPCA}\vspace{-.05in}
 RPCA may be formulated as the following  non-convex optimization problem
\begin{multline} \label{eq:optimRPCA}
    \argmin_{\L, \S \in \RR^{d_1 \times d_2}} { \| \M^\star - \L - \S \|_{\mathrm{F}}}, \\
    \text{subject to } \rank(\L) \leq r \text{ and } \|\S\|_0 \leq |\Omega|,
\end{multline}
where  $\|\cdot\|_{\mathrm{F}}$ is the Frobenius norm, $r \geq \rank (\L^\star)$ upper bounds the rank of the   low rank matrix $\L^\star$, and $k \geq |\Omega|$ upper bounds the  cardinally of the support of the  sparse matrix $\S^\star$.

Netrapalli {\em et al.}~\cite{altproj:2014} proposed to solve \eqref{eq:optimRPCA} using the alternating projections (AltProj) method, which projects $\M^\star - \S_k$ onto the space of low rank matrices and $\M^\star - \L_k$ onto the set of sparse matrices in an alternating manner at each iteration~$k$. It enjoys a computational complexity of $\mathcal{O}(d_1 d_2 r^2)$ per iteration. Building upon AltProj,  Cai {\em et al.}~\cite{accaltproj:2019} proposed an accelerated version known as AccAltProj  with an improved complexity $\mathcal{O}(d_1 d_2 r)$. Later, Cai {\em et al.}~\cite{ircur:2021} introduced the Iterated Robust CUR (IRCUR), which is a variant of AltProj with a per-iteration complexity of $\mathcal{O}(r^2 n \log n )$, where $n=\max(d_1, d_2)$. This is achieved by operating on submatrices, hence avoiding expensive computations on full matrices. However, it is widely acknowledged that CUR-based decompositions are less accurate than SVD-based ones. 

We briefly describe  AccAltProj in Alg.~\ref{algo accaltproj}, before we move on to the proposed unrolled model. Here, $\mathcal{M}_r$ denotes the set of rank-$r$ matrices, and $T_k$ denotes the tangent space of $\mathcal{M}_r$ at $\L_k$. The $i$-th largest singular value of a matrix $\mathbf{X}$ is denoted as $\sigma_i^{(\mathbf{X})}$. 
The operator $\H_r$ represents the truncated SVD operation at rank $r$ and $\T_\zeta$ represents the hard-thresholding operator (i.e., the proximity operator of $\ell_0$) with threshold $\zeta$.

AccAltProj differs from AltProj by performing a tangent space projection on $T_k$ rather than directly projecting $\M^\star - \S_k$ onto $\mathcal{M}_k$. This is  followed by projecting the intermediate matrix onto $\mathcal{M}_r$ to obtain $\L_{k+1}$ before projecting $\M^\star - \L_{k+1}$ back onto the set of sparse matrices.
Cai {\em et al.}~\cite{accaltproj:2019} derived the projection operator onto $T_k$ as: 
 \begin{align}  
     P_{T_k}(\mathbf{A}) 
     = \begin{bmatrix} \U_k & \Q_1 \end{bmatrix} 
     \begin{bmatrix} \U_k^\top \mathbf{A} \V_k & \R_2^\top\\ \R_1 & \mathbf{0} \end{bmatrix} 
     \begin{bmatrix} \V_k^\top \\ \Q_2^\top \end{bmatrix},  
 \end{align}
where $\U_k$, $\V_k$ contain the singular vectors from the truncated SVD of $\L_k = \U_k \bm{\Sigma}_k \V_k^\top$, and $(\Q_1,\R_1)$ and $(\Q_2,\R_2)$ are the factors from the  QR decompositions of $(\mathbf{I} - \V_k\V_k^\top)(\M^\star - \S_k)^\top\U_k$ and $(\mathbf{I} - \U_k\U_k^\top)(\M^\star - \S_k)\V_k$ respectively. \vspace{-.05in}

\begin{algorithm}[t]
    \caption{Accelerated Alternating Projections (AccAltProj)} \label{algo accaltproj}
    \KwIn{$\M^\star$, $r$, $\epsilon$, $\beta_{\mathrm{init}}$, $\beta$, $\gamma$}
    \nl $\zeta_{-1} \gets \beta_{\mathrm{init}} \cdot \sigma_1^{(\M^\star)}$\;
    \nl $\S_{-1} \gets \T_{\zeta_{-1}}(\M^\star)$\;
    \nl $\L_0 \gets \H_r(\M^\star - \S_{-1})$\;
    \nl $\zeta_0 \gets \beta \cdot \sigma_1^{(\M^\star - \S_{-1})}$\;
    \nl $\S_0 \gets \T_{\zeta_0}(\M^\star - \L_0)$\;
    \nl $k \gets 0$\;
    \nl \While{$\|\M^\star - \L_k - \S_k\|_{\mathrm{F}}/ \|\M^\star\|_{\mathrm{F}} \geq \epsilon$}{
    \nl     $\mathbf{P}_{k+1} \gets P_{T_k}(\M^\star - \S_k)$\;
    \nl     $\L_{k+1} \gets \H_r(\mathbf{P}_{k+1})$\;
    \nl     $\zeta_{k+1} \gets \beta (\sigma_{r+1}^{(\mathbf{P}_{k+1})} + \gamma _{k+1}\sigma_1^{(\mathbf{P}_{k+1})})$\;
    \nl     $\S_{k+1} \gets \T_{\zeta_{k+1}}(\M^\star - \L_{k+1})$\;
    \nl     $k \gets k + 1$\;
    }
    \KwOut{$\L_k, \S_k$}  
\end{algorithm}
\vspace{-.05in}

\vspace{-.05in}
\subsection{Deep Unrolling}
\vspace{-.05in}

A tedious task in implementing iterative optimization algorithms is to tune their   hyperparameters  (e.g.,   stepsize,  regularisation parameters). To circumvent this problem, {\em unrolled} versions of standard algorithms \cite{Monga2021} have recently been developed. 
In essence, algorithm unrolling or unfolding consists in converting an iterative algorithm into a neural network. One iteration of the iterative algorithm is being transformed to one layer of the neural network. The benefits of this approach include neural network interpretability and automatic parameter learning. 
Following this line of thought, we propose to unroll the Accelerated Alternating Projections algorithm (Alg.~\ref{algo accaltproj}).

\section{Proposed unrolled A\MakeLowercase{cc}A\MakeLowercase{lt}P\MakeLowercase{roj}}
\label{sec:prop_algo}

We adopt AccAltProj as our baseline model to unroll as it is fast compared to most existing algorithms and is more robust compared to IRCUR. We follow the idea from the Learned Iterative Soft Thresholding Algorithm \cite{lista:2010} to design a non-linear feed-forward architecture with  a fixed number of layers.

As $\beta$ and $\gamma$ are fixed heuristically in AccAltProj, we chose to learn them in the unrolled network. The parameter $\beta$ controls the variance of matrix elements of recovered $\hat{\L}$ while $\gamma$ controls the rate of convergence \cite{altproj:2014}. They also play a key role for the theoretical  guarantee of AccAltProj. More precisely, if properly chosen, the initial guesses $\mathbf{S}_{-1}$ and $\mathbf{L}_0$ generated at Lines 1 to 5 of Alg.~\ref{algo accaltproj}  fulfill the required condition for local convergence of AccAltProj~\cite[Theorem~1]{accaltproj:2019}. Learning $\beta$ and $\gamma$ automatically  allows our model to be customisable to  use cases where datasets share similar properties for the underlying low-rank and sparse components.


\subsection{Using the Minimax Concave Penalty (MCP) instead of the $\ell_0$ or $\ell_1$ norms}

One challenge when developing an unrolled version of AccAltProj is that we are unable to directly use hard-thresholding  for the non-linear activation. This is because it is not subdifferentiable, a property needed to deploy gradient-based optimizers to learn the parameters $\beta$ and $\gamma$~\cite{lista:2010}.

LRPCA \cite{lrpca:2021} tackled this problem by replacing the hard-thresholding operator  with the soft-thresholding operator  in their unrolled model. 
However, the soft-thresholding operator  is the proximal mapping of the $\ell_1$ norm while the hard-thresholding operator  is the proximal mapping of the $\ell_0$ pseudo-norm. As such, vanilla soft-thresholding is not  suitable for our objective in~\eqref{eq:optimRPCA}.

Taking the best of both worlds, we consider in this work the {\em  Minimax Concave Penalty} (MCP) \cite{mcp:2010}, defined as 
\begin{eqnarray}\label{eq:MCP}
    \mathrm{MCP}(x; \zeta, \upsilon) =
    \begin{cases}
        \frac{\upsilon \zeta^2}{2},& \text{if } |x| > \upsilon \zeta \\
    \zeta|x|-\frac{x^2}{2\upsilon} ,              & \text{if } |x| \leq \upsilon \zeta, 
    \end{cases}
\end{eqnarray}
where $\zeta$ is the threshold, and $\upsilon >1$ is a parameter controlling the concavity of the penalty. It has a close relationship to  the $\ell_0$ pseudo-norm, from both the statistical~\cite{mcp:2010} and optimization viewpoints~\cite{unifiedl2l0}, while being  subdifferentiable.

\begin{figure}[htb]
\begin{minipage}[b]{.48\linewidth}
  \centering
  \centerline{\includegraphics[width=4.2cm]{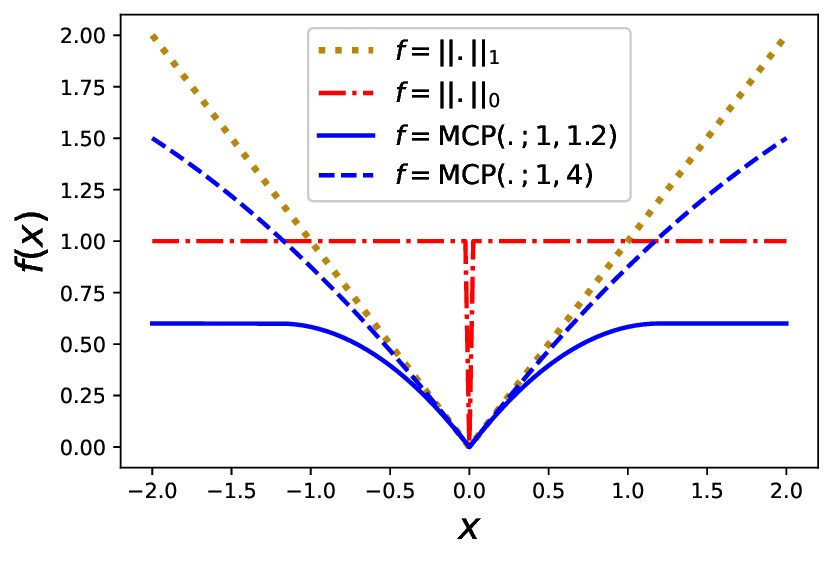}}
\end{minipage}
\hfill
\begin{minipage}[b]{0.48\linewidth}
  \centering
  \centerline{\includegraphics[width=4.2cm]{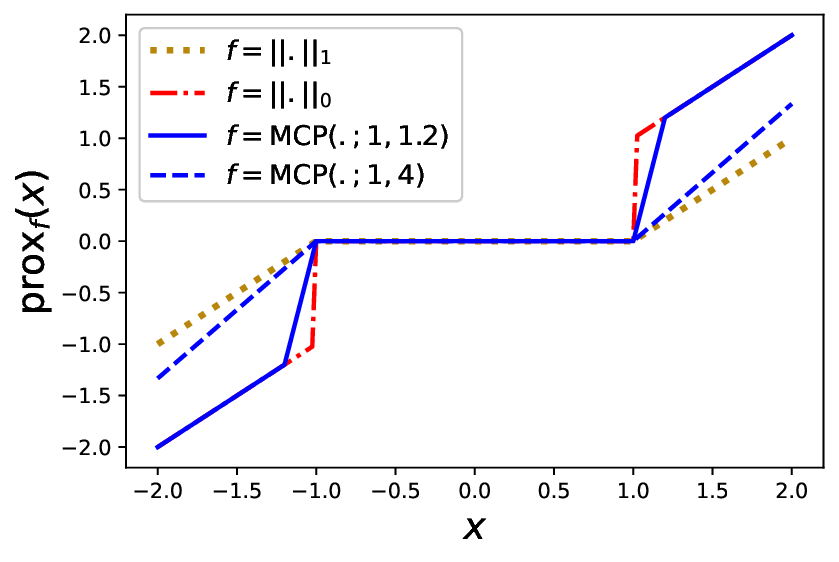}}
\end{minipage}
\caption{Penalty functions and  their proximal mappings.}
\label{fig:res}
\end{figure}

Its proximal mapping $\mathcal{P}(x; \zeta, \upsilon) := \mathrm{prox}_{\mathrm{MCP}}(x; \zeta, \upsilon)$ is\begin{align} \label{eqn 7}
\P(x; \zeta, \upsilon) = 
    \mathrm{sign}(x)\min \left\{ \frac{\upsilon \max(|x| - \zeta, 0)}{\upsilon - 1} , |x| \right\}
\end{align}
which corresponds to the ``firm thresholding operator'' \cite{firmthres:2016}, a compromise between soft- and hard- thresholding. In the unrolled version of Alg.~\ref{algo accaltproj}, we  use $\P(\cdot; \zeta, \upsilon)$ in place of the hard thresholding operator $\T_{\zeta}$. The penalty functions and their corresponding proximal mappings are shown in Fig.~\ref{fig:res}.

\vspace{-.05in}
\subsection{Unrolled AccAltProj}
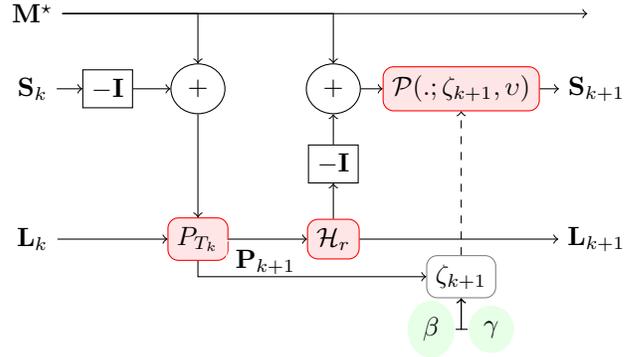
\begin{figure}[!ht]
\centering
\begin{tikzpicture}
\node[draw=white] (In_M) at (0,0) {$\M^\star$};
\node[draw=white] (In_S) at (0,-1) {$\S_k$};
\node[draw=white] (In_L) at (0,-3) {$\L_k$};
\node[draw] (Id1) at (1,-1) {$-\I$};
\node[draw,ellipse] (Plus1) at (2.2,-1) {$+$};
\node[draw=red,fill=red!10,rounded corners] (P) at (2.2,-3) {$P_{T_k}$};
\node[draw=white]  at (3.1,-3.3) {$\mathbf{P}_{k+1}$};
\node[draw=red,fill=red!10,rounded corners] (H) at (4,-3) {$\H_{r}$};
\node[draw] (Id2) at (4,-2) {$-\I$};
\node[draw,ellipse] (Plus2) at (4,-1) {$+$};
\node[draw=red,fill=red!10,rounded corners] (mP) at (5.7,-1) {$\P(.; \zeta_{k+1}, \upsilon)$};

\node[draw=gray,rounded corners] (zet) at (5.7,-3.5) {$\zeta_{k+1}$};

\node[draw=white,fill=green!10,ellipse] (b) at (5.3,-4.2) {$\beta$};
\node[draw=white,fill=green!10,ellipse] (g) at (6.1,-4.2) {$\gamma$};

\node[draw=white] (blank) at (7.5,0) {};
\node[draw=white] (Out_S) at (7.5,-1) {$\S_{k+1}$};
\node[draw=white] (Out_L) at (7.5,-3) {$\L_{k+1}$};

\draw[black,->] (P) |- (zet);
\draw[black,->] (b) -| (zet);
\draw[black,->] (g) -| (zet);
\draw[black,->,dashed] (zet) -- (mP);

\draw[black,->] (In_M) -- (blank);
\draw[black,->] (In_M) -| (Plus1);
\draw[black,->] (In_M) -| (Plus2);

\draw[black,->] (In_S) -- (Id1);
\draw[black,->] (Id1) -- (Plus1);
\draw[black,->] (Plus2) -- (mP);
\draw[black,->] (mP) -- (Out_S);

\draw[black,->] (In_L) -- (P);
\draw[black,->] (Plus1) -- (P);
\draw[black,->] (P) -- (H);
\draw[black,->] (H) -- (Id2);
\draw[black,->] (Id2) -- (Plus2);
\draw[black,->] (H) -- (Out_L);

\end{tikzpicture}
\caption{One layer of the unrolled RPCA algorithm.  Note that $\zeta_{k+1}$ is a function of $(\beta,\gamma)$, defined in Line 10 of Alg.~\ref{algo accaltproj}. \label{fig:layer}}
\end{figure}

\vspace{-.05in}
We consider  an unrolled version of the Modified Accelerated Alternating Projections and  refer to it as the unrolled RPCA algorithm. Each iteration is thus transformed in one layer as shown in Fig.~\ref{fig:layer}. We use this neural network to learn   $\beta$ and $\gamma$ while keeping $\upsilon$ fixed ($\upsilon = 1.05$).

\vspace{-.05in}
\subsection{Training Criteria}
\label{ssec:train_crit}
\vspace{-.05in}

While it is possible to use adaptive parameters (i.e., separate $\beta_k$, $\gamma_k$ for each layer $k$), we choose to learn only a single  $(\beta,\gamma)$ that is {\em shared} across the layers. In the unrolled model, we initialise $\beta=\frac{1}{2 \cdot \sqrt[4]{d_1 \times d_2}}$ and $\gamma=0.7$ since they are the default values used in AccAltProj \cite{accaltproj:2019}.

Consider a set of input data $\{\M_\mathrm{train}^q\}_{q=1}^Q$ and the associated sparse and low-rank decomposition that we denote by $\{(\L_\mathrm{train}^q, \S_\mathrm{train}^q)\}_{q=1}^Q$. These can be obtained either via simulations or through the application of classical RPCA iterative algorithms on $\M_\mathrm{train}^q$. Then, following~\cite{corona:2020}, we learn the two parameters $\gamma$ and $\beta$ via 
\begin{align}
    (\hat{\gamma},\hat{\beta}) \in   \argmin_{(\gamma,\beta) \in \RR^2}&  \, \sum_{q=1}^Q \mathcal{L}(\L_\mathrm{train}^q,\L^q) + \mathcal{L}(\S_\mathrm{train}^q,\S^q) \\
     \text{subject to} & \; (\L^q, \S^q) = \mathcal{N}(\gamma,\beta; \M_\mathrm{train}^q) \notag
\end{align}
where $\mathcal{N}$ is defined by cascading layers as in Fig.~\ref{fig:layer} (unrolled network). Finally, we set the loss  $\mathcal{L}$ to be the relative error $\mathcal{L}(\tilde{\mathbf{X}},\mathbf{X}) = {\|\tilde{\mathbf{X}} - \mathbf{X}\|_{\mathrm{F}}^2}/{\|\mathbf{X}\|_{\mathrm{F}}^2}$.


\vspace{-.1in}
\section{Numerical experiments}
\label{sec:simu}
\vspace{-.1in}
To illustrate the effectiveness of the proposed approach, we performed experiments on two settings: a fully controlled one through synthetic simulations and a realistic one in the context of face modelling. The code to reproduce the simulations will be released. 


\subsection{Simulated/Sythetic Data}


{\bf Problem setup:} The synthetic data are generated as in~\cite{accaltproj:2019}, i.e., let $\L^\star = \mathbf{UV}^\top$, where $\U,\V \in \RR^{d \times r}$ contain elements generated i.i.d.\ from the standard normal distribution. Similarly, the components of $\S^\star$ are  sampled i.i.d.\ and uniformly from the interval $[-c \cdot \mathbb{E}(| [\L^\star]_{ij} |), \ c \cdot \mathbb{E}(| [\L^\star]_{ij} |)]$ where $c>0$. The positions of the non-zero elements are randomly sampled without replacement. In the following, the matrix $\S^\star$ is said to be {\em $\alpha$-sparse} if each of its rows and columns contain at most $\alpha d$ non-zero elements. Finally, given a generated pair $(\L^\star,\S^\star)$, we generate an input-target training data matrix as  $\M_{\mathrm{train}} = \M^\star = \L^\star + \S^\star$ and  ($\L_{\mathrm{train}},\S_{\mathrm{train}}$) is obtained via IRCUR applied on  $\M_{\mathrm{train}}$.

For this experiment, we fix the dimensions to $d_1=d_2=d  = 250$ and the rank to $r=2$. We consider several simulated data sets generated by varying the sparsity level (controlled by $\alpha$) and the amplitude  (controlled by $c$) of the sparse component~$\S^\star$. More precisely, we consider the following cases to assess  the performance of our unrolled network:

\begin{table}[!ht]
\begin{center}
\begin{tabular}{ccccc}
 \hline
   & Case 1 & Case 2 & Case 3 & Case 4 \\
 \hline
($\alpha,c$) & ($0.1,1$)  & ($0.3,1$)  & ($0.01,1$)  & ($0.1,10$) \\
\hline
\end{tabular}
\caption{Table showing different experimental settings. \label{tab:scena}}
\end{center}
\end{table}

For each case, we generate a total of $300$ samples, and split them into $180$ training samples and $120$ test samples. The unrolled network  is trained for a total of $8$ epochs.

The metrics that we use to quantify the performances of the unrolled model and its competitors are as follows:
\begin{align}
    \epsilon_{\mathrm{L}}(\L_{\mathrm{out}}) &:= \| \L^\star - \L_{\mathrm{out}} \|_{\mathrm{F}}, \label{eq:metric fro-L}\\
    \epsilon_{\mathrm{S}}(\S_{\mathrm{out}}) &:= \| \S^\star - \S_{\mathrm{out}} \|_{\mathrm{F}}, \label{eq:metric fro-S}\\
    \epsilon_{\mathrm{M}}(\L_{\mathrm{out}}, \S_{\mathrm{out}}) &:= \| \M^\star - \L_{\mathrm{out}} - \S_{\mathrm{out}} \|_{\mathrm{F}} / \| \M^\star \|_{\mathrm{F}}, \label{eq:metric fro-M}\\
    \epsilon_{\mathrm{supp}}(\S_{\mathrm{out}}) &:= \frac{1}{d^2} (\mathbbm{1}_{ \{ [\S^\star]_{ij} = 0, [\S_{\mathrm{out}}]_{ij} \neq 0\} } \nonumber \\ &\qquad\qquad + \mathbbm{1}_{ \{ [\S^\star]_{ij} \neq 0, [\S_{\mathrm{out}}]_{ij} = 0\} }), \label{eq:metric supp-S}
\end{align}
where $\L_{\mathrm{out}}$, $\S_{\mathrm{out}}$ are placeholders for the outputs that could be computed from IRCUR, AccAltProj, or the unrolled model (after training). These four errors respectively quantify the accuracies on 1) the estimation of $\L^\star$, 2) the estimation of $\S^\star$, 3) the overall matrix $\M^\star$  and 4)   support recovery of $\S^\star$.

\begin{figure*}
 \pgfplotsset{every tick label/.append style={font=\scriptsize}}
    \centering
    \begin{tikzpicture}
		\begin{groupplot}[group style={group size= 2 by 2,                      
    					  horizontal sep=0.9cm, vertical sep=1.2cm},     
					      grid=both,                         
    				 width=0.5\textwidth,height=0.3\textwidth,
    				 ymode=log,legend style={at={(1.55,-0.1)}},
                      legend columns=4,
    				 ylabel shift = -2 pt,xlabel shift = -2 pt,title style={yshift=-5pt}] 
			\nextgroupplot[title={$\epsilon_{\mathrm{M}}$},xmin=1,xmax=4,xtick={1,2,3,4}, ymin=0.00000001,ymax=0.000001,xticklabels={,,}] 
			\pgfplotstableread{images/eM_IRCUR_mean.dat}{\tableMean}
			\pgfplotstableread{images/eM_IRCUR_std.dat}{\tableStd}
			\pgfplotstableset{
     			create on use/x/.style={create col/copy column from table={\tableMean}{x}}, 
     			create on use/y1/.style={create col/copy column from table={\tableMean}{y}},
     			create on use/y2/.style={create col/copy column from table={\tableStd}{y}}, 
     			create on use/sum/.style={create col/expr={\thisrow{y1}+\thisrow{y2}}},     
     			create on use/diff/.style={create col/expr={\thisrow{y1}-\thisrow{y2}}}
			}
			\pgfplotstablenew[columns={x,y1,y2,sum}]{\pgfplotstablegetrowsof{\tableMean}}\tableStdp
			\pgfplotstablenew[columns={x,y1,y2,diff}]{\pgfplotstablegetrowsof{\tableMean}}\tableStdm
			\addplot[darkgreen,very thick,mark=diamond,mark size=3pt] table{\tableMean};
			\addplot[name path=A,darkgreen,dashed,forget plot] table [x=x, y=sum]{\tableStdp};
			\addplot[name path=B,darkgreen,dashed,forget plot] table [x=x, y=diff]{\tableStdm};
			\addplot[darkgreen!20,forget plot,opacity=0.5] fill between[of=A and B];
			\pgfplotstableread{images/eM_AccAltProj_mean.dat}{\tableMean}
			\pgfplotstableread{images/eM_AccAltProj_std.dat}{\tableStd}
			\pgfplotstableset{
     			create on use/x/.style={create col/copy column from table={\tableMean}{x}}, 
     			create on use/y1/.style={create col/copy column from table={\tableMean}{y}},
     			create on use/y2/.style={create col/copy column from table={\tableStd}{y}}, 
     			create on use/sum/.style={create col/expr={\thisrow{y1}+\thisrow{y2}}},     
     			create on use/diff/.style={create col/expr={\thisrow{y1}-\thisrow{y2}}}
			}
			\pgfplotstablenew[columns={x,y1,y2,sum}]{\pgfplotstablegetrowsof{\tableMean}}\tableStdp
			\pgfplotstablenew[columns={x,y1,y2,diff}]{\pgfplotstablegetrowsof{\tableMean}}\tableStdm
			\addplot[darkred,very thick,mark=square] table{\tableMean};
			\addplot[name path=A,darkred,dashed,forget plot] table [x=x, y=sum]{\tableStdp};
			\addplot[name path=B,darkred,dashed,forget plot] table [x=x, y=diff]{\tableStdm};
			\addplot[darkred!20,forget plot,opacity=0.5] fill between[of=A and B]; 
			\pgfplotstableread{images/eM_Unrolled_mean.dat}{\tableMean}
			\pgfplotstableread{images/eM_Unrolled_std.dat}{\tableStd}
			\pgfplotstableset{
     			create on use/x/.style={create col/copy column from table={\tableMean}{x}}, 
     			create on use/y1/.style={create col/copy column from table={\tableMean}{y}},
     			create on use/y2/.style={create col/copy column from table={\tableStd}{y}}, 
     			create on use/sum/.style={create col/expr={\thisrow{y1}+\thisrow{y2}}},     
     			create on use/diff/.style={create col/expr={\thisrow{y1}-\thisrow{y2}}}
			}
			\pgfplotstablenew[columns={x,y1,y2,sum}]{\pgfplotstablegetrowsof{\tableMean}}\tableStdp
			\pgfplotstablenew[columns={x,y1,y2,diff}]{\pgfplotstablegetrowsof{\tableMean}}\tableStdm
			\addplot[blue,very thick,mark=o] table{\tableMean};
			\addplot[name path=A,blue,dashed,forget plot] table [x=x, y=sum]{\tableStdp};
			\addplot[name path=B,blue,dashed,forget plot] table [x=x, y=diff]{\tableStdm};
			\addplot[blue!20,forget plot,opacity=0.5] fill between[of=A and B]; 
			\pgfplotstableread{images/eM_Unrolled_st_mean.dat}{\tableMean}
			\pgfplotstableread{images/eM_Unrolled_st_std.dat}{\tableStd}
			\pgfplotstableset{
     			create on use/x/.style={create col/copy column from table={\tableMean}{x}}, 
     			create on use/y1/.style={create col/copy column from table={\tableMean}{y}},
     			create on use/y2/.style={create col/copy column from table={\tableStd}{y}}, 
     			create on use/sum/.style={create col/expr={\thisrow{y1}+\thisrow{y2}}},     
     			create on use/diff/.style={create col/expr={\thisrow{y1}-\thisrow{y2}}}
			}
			\pgfplotstablenew[columns={x,y1,y2,sum}]{\pgfplotstablegetrowsof{\tableMean}}\tableStdp
			\pgfplotstablenew[columns={x,y1,y2,diff}]{\pgfplotstablegetrowsof{\tableMean}}\tableStdm
			\addplot[gray,very thick,mark=triangle] table{\tableMean};
			\addplot[name path=A,gray,dashed,forget plot] table [x=x, y=sum]{\tableStdp};
			\addplot[name path=B,gray,dashed,forget plot] table [x=x, y=diff]{\tableStdm};
			\addplot[gray!20,forget plot,opacity=0.5] fill between[of=A and B]; 
            \legend{{\scriptsize IRCUR},{\scriptsize AccAltProj},{\scriptsize Unrolled},{\scriptsize Unrolled (st)}};
		\nextgroupplot[title={$\epsilon_{\mathrm{L}}$},xmin=1,xmax=4,xtick={1,2,3,4}, ymin=0.00005,ymax=0.0005,xticklabels={,,}] 
  \pgfplotstableread{images/eL_IRCUR_mean.dat}{\tableMean}
			\pgfplotstableread{images/eL_IRCUR_std.dat}{\tableStd}
			\pgfplotstableset{
     			create on use/x/.style={create col/copy column from table={\tableMean}{x}}, 
     			create on use/y1/.style={create col/copy column from table={\tableMean}{y}},
     			create on use/y2/.style={create col/copy column from table={\tableStd}{y}}, 
     			create on use/sum/.style={create col/expr={\thisrow{y1}+\thisrow{y2}}},     
     			create on use/diff/.style={create col/expr={\thisrow{y1}-\thisrow{y2}}}
			}
			\pgfplotstablenew[columns={x,y1,y2,sum}]{\pgfplotstablegetrowsof{\tableMean}}\tableStdp
			\pgfplotstablenew[columns={x,y1,y2,diff}]{\pgfplotstablegetrowsof{\tableMean}}\tableStdm
			\addplot[darkgreen,very thick,mark=diamond,mark size=3pt] table{\tableMean};
			\addplot[name path=A,darkgreen,dashed,forget plot] table [x=x, y=sum]{\tableStdp};
			\addplot[name path=B,darkgreen,dashed,forget plot] table [x=x, y=diff]{\tableStdm};
			\addplot[darkgreen!20,forget plot,opacity=0.5] fill between[of=A and B];
			\pgfplotstableread{images/eL_AccAltProj_mean.dat}{\tableMean}
			\pgfplotstableread{images/eL_AccAltProj_std.dat}{\tableStd}
			\pgfplotstableset{
     			create on use/x/.style={create col/copy column from table={\tableMean}{x}}, 
     			create on use/y1/.style={create col/copy column from table={\tableMean}{y}},
     			create on use/y2/.style={create col/copy column from table={\tableStd}{y}}, 
     			create on use/sum/.style={create col/expr={\thisrow{y1}+\thisrow{y2}}},     
     			create on use/diff/.style={create col/expr={\thisrow{y1}-\thisrow{y2}}}
			}
			\pgfplotstablenew[columns={x,y1,y2,sum}]{\pgfplotstablegetrowsof{\tableMean}}\tableStdp
			\pgfplotstablenew[columns={x,y1,y2,diff}]{\pgfplotstablegetrowsof{\tableMean}}\tableStdm
			\addplot[darkred,very thick,mark=square] table{\tableMean};
			\addplot[name path=A,darkred,dashed,forget plot] table [x=x, y=sum]{\tableStdp};
			\addplot[name path=B,darkred,dashed,forget plot] table [x=x, y=diff]{\tableStdm};
			\addplot[darkred!20,forget plot,opacity=0.5] fill between[of=A and B]; 
			\pgfplotstableread{images/eL_Unrolled_mean.dat}{\tableMean}
			\pgfplotstableread{images/eL_Unrolled_std.dat}{\tableStd}
			\pgfplotstableset{
     			create on use/x/.style={create col/copy column from table={\tableMean}{x}}, 
     			create on use/y1/.style={create col/copy column from table={\tableMean}{y}},
     			create on use/y2/.style={create col/copy column from table={\tableStd}{y}}, 
     			create on use/sum/.style={create col/expr={\thisrow{y1}+\thisrow{y2}}},     
     			create on use/diff/.style={create col/expr={\thisrow{y1}-\thisrow{y2}}}
			}
			\pgfplotstablenew[columns={x,y1,y2,sum}]{\pgfplotstablegetrowsof{\tableMean}}\tableStdp
			\pgfplotstablenew[columns={x,y1,y2,diff}]{\pgfplotstablegetrowsof{\tableMean}}\tableStdm
			\addplot[blue,very thick,mark=o] table{\tableMean};
			\addplot[name path=A,blue,dashed,forget plot] table [x=x, y=sum]{\tableStdp};
			\addplot[name path=B,blue,dashed,forget plot] table [x=x, y=diff]{\tableStdm};
			\addplot[blue!20,forget plot,opacity=0.5] fill between[of=A and B]; 
			\pgfplotstableread{images/eL_Unrolled_st_mean.dat}{\tableMean}
			\pgfplotstableread{images/eL_Unrolled_st_std.dat}{\tableStd}
			\pgfplotstableset{
     			create on use/x/.style={create col/copy column from table={\tableMean}{x}}, 
     			create on use/y1/.style={create col/copy column from table={\tableMean}{y}},
     			create on use/y2/.style={create col/copy column from table={\tableStd}{y}}, 
     			create on use/sum/.style={create col/expr={\thisrow{y1}+\thisrow{y2}}},     
     			create on use/diff/.style={create col/expr={\thisrow{y1}-\thisrow{y2}}}
			}
			\pgfplotstablenew[columns={x,y1,y2,sum}]{\pgfplotstablegetrowsof{\tableMean}}\tableStdp
			\pgfplotstablenew[columns={x,y1,y2,diff}]{\pgfplotstablegetrowsof{\tableMean}}\tableStdm
			\addplot[gray,very thick,mark=triangle] table{\tableMean};
			\addplot[name path=A,gray,dashed,forget plot] table [x=x, y=sum]{\tableStdp};
			\addplot[name path=B,gray,dashed,forget plot] table [x=x, y=diff]{\tableStdm};
			\addplot[gray!20,forget plot,opacity=0.5] fill between[of=A and B]; 
		\nextgroupplot[title={$\epsilon_{\mathrm{S}}$},xlabel={Cases},xmin=1,xmax=4,xtick={1,2,3,4}, ymin=0.000005,ymax=0.02]
  \pgfplotstableread{images/eS_IRCUR_mean.dat}{\tableMean}
			\pgfplotstableread{images/eS_IRCUR_std.dat}{\tableStd}
			\pgfplotstableset{
     			create on use/x/.style={create col/copy column from table={\tableMean}{x}}, 
     			create on use/y1/.style={create col/copy column from table={\tableMean}{y}},
     			create on use/y2/.style={create col/copy column from table={\tableStd}{y}}, 
     			create on use/sum/.style={create col/expr={\thisrow{y1}+\thisrow{y2}}},     
     			create on use/diff/.style={create col/expr={\thisrow{y1}-\thisrow{y2}}}
			}
			\pgfplotstablenew[columns={x,y1,y2,sum}]{\pgfplotstablegetrowsof{\tableMean}}\tableStdp
			\pgfplotstablenew[columns={x,y1,y2,diff}]{\pgfplotstablegetrowsof{\tableMean}}\tableStdm
			\addplot[darkgreen,very thick,mark=diamond,mark size=3pt] table{\tableMean};
			\addplot[name path=A,darkgreen,dashed,forget plot] table [x=x, y=sum]{\tableStdp};
			\addplot[name path=B,darkgreen,dashed,forget plot] table [x=x, y=diff]{\tableStdm};
			\addplot[darkgreen!20,forget plot,opacity=0.5] fill between[of=A and B];
			\pgfplotstableread{images/eS_AccAltProj_mean.dat}{\tableMean}
			\pgfplotstableread{images/eS_AccAltProj_std.dat}{\tableStd}
			\pgfplotstableset{
     			create on use/x/.style={create col/copy column from table={\tableMean}{x}}, 
     			create on use/y1/.style={create col/copy column from table={\tableMean}{y}},
     			create on use/y2/.style={create col/copy column from table={\tableStd}{y}}, 
     			create on use/sum/.style={create col/expr={\thisrow{y1}+\thisrow{y2}}},     
     			create on use/diff/.style={create col/expr={\thisrow{y1}-\thisrow{y2}}}
			}
			\pgfplotstablenew[columns={x,y1,y2,sum}]{\pgfplotstablegetrowsof{\tableMean}}\tableStdp
			\pgfplotstablenew[columns={x,y1,y2,diff}]{\pgfplotstablegetrowsof{\tableMean}}\tableStdm
			\addplot[darkred,very thick,mark=square] table{\tableMean};
			\addplot[name path=A,darkred,dashed,forget plot] table [x=x, y=sum]{\tableStdp};
			\addplot[name path=B,darkred,dashed,forget plot] table [x=x, y=diff]{\tableStdm};
			\addplot[darkred!20,forget plot,opacity=0.5] fill between[of=A and B]; 
			\pgfplotstableread{images/eS_Unrolled_mean.dat}{\tableMean}
			\pgfplotstableread{images/eS_Unrolled_std.dat}{\tableStd}
			\pgfplotstableset{
     			create on use/x/.style={create col/copy column from table={\tableMean}{x}}, 
     			create on use/y1/.style={create col/copy column from table={\tableMean}{y}},
     			create on use/y2/.style={create col/copy column from table={\tableStd}{y}}, 
     			create on use/sum/.style={create col/expr={\thisrow{y1}+\thisrow{y2}}},     
     			create on use/diff/.style={create col/expr={\thisrow{y1}-\thisrow{y2}}}
			}
			\pgfplotstablenew[columns={x,y1,y2,sum}]{\pgfplotstablegetrowsof{\tableMean}}\tableStdp
			\pgfplotstablenew[columns={x,y1,y2,diff}]{\pgfplotstablegetrowsof{\tableMean}}\tableStdm
			\addplot[blue,very thick,mark=o] table{\tableMean};
			\addplot[name path=A,blue,dashed,forget plot] table [x=x, y=sum]{\tableStdp};
			\addplot[name path=B,blue,dashed,forget plot] table [x=x, y=diff]{\tableStdm};
			\addplot[blue!20,forget plot,opacity=0.5] fill between[of=A and B]; 
			\pgfplotstableread{images/eS_Unrolled_st_mean.dat}{\tableMean}
			\pgfplotstableread{images/eS_Unrolled_st_std.dat}{\tableStd}
			\pgfplotstableset{
     			create on use/x/.style={create col/copy column from table={\tableMean}{x}}, 
     			create on use/y1/.style={create col/copy column from table={\tableMean}{y}},
     			create on use/y2/.style={create col/copy column from table={\tableStd}{y}}, 
     			create on use/sum/.style={create col/expr={\thisrow{y1}+\thisrow{y2}}},     
     			create on use/diff/.style={create col/expr={\thisrow{y1}-\thisrow{y2}}}
			}
			\pgfplotstablenew[columns={x,y1,y2,sum}]{\pgfplotstablegetrowsof{\tableMean}}\tableStdp
			\pgfplotstablenew[columns={x,y1,y2,diff}]{\pgfplotstablegetrowsof{\tableMean}}\tableStdm
			\addplot[gray,very thick,mark=triangle] table{\tableMean};
			\addplot[name path=A,gray,dashed,forget plot] table [x=x, y=sum]{\tableStdp};
			\addplot[name path=B,gray,dashed,forget plot] table [x=x, y=diff]{\tableStdm};
			\addplot[gray!20,forget plot,opacity=0.5] fill between[of=A and B]; 
	      \nextgroupplot[title={$\epsilon_{\textrm{supp}}$},xlabel={Cases},xmin=1,xmax=4,xtick={1,2,3,4}, ymin=0.0000001,ymax=0.55] 
       \pgfplotstableread{images/esupp_IRCUR_mean.dat}{\tableMean}
			\pgfplotstableread{images/esupp_IRCUR_std.dat}{\tableStd}
			\pgfplotstableset{
     			create on use/x/.style={create col/copy column from table={\tableMean}{x}}, 
     			create on use/y1/.style={create col/copy column from table={\tableMean}{y}},
     			create on use/y2/.style={create col/copy column from table={\tableStd}{y}}, 
     			create on use/sum/.style={create col/expr={\thisrow{y1}+\thisrow{y2}}},     
     			create on use/diff/.style={create col/expr={\thisrow{y1}-\thisrow{y2}}}
			}
			\pgfplotstablenew[columns={x,y1,y2,sum}]{\pgfplotstablegetrowsof{\tableMean}}\tableStdp
			\pgfplotstablenew[columns={x,y1,y2,diff}]{\pgfplotstablegetrowsof{\tableMean}}\tableStdm
			\addplot[darkgreen,very thick,mark=diamond,mark size=3pt] table{\tableMean};
			\addplot[name path=A,darkgreen,dashed,forget plot] table [x=x, y=sum]{\tableStdp};
			\addplot[name path=B,darkgreen,dashed,forget plot] table [x=x, y=diff]{\tableStdm};
			\addplot[darkgreen!20,forget plot,opacity=0.5] fill between[of=A and B];
			\pgfplotstableread{images/esupp_Unrolled_st_mean.dat}{\tableMean}
			\pgfplotstableread{images/esupp_Unrolled_st_std.dat}{\tableStd}
			\pgfplotstableset{
     			create on use/x/.style={create col/copy column from table={\tableMean}{x}}, 
     			create on use/y1/.style={create col/copy column from table={\tableMean}{y}},
     			create on use/y2/.style={create col/copy column from table={\tableStd}{y}}, 
     			create on use/sum/.style={create col/expr={\thisrow{y1}+\thisrow{y2}}},     
     			create on use/diff/.style={create col/expr={\thisrow{y1}-\thisrow{y2}}}
			}
			\pgfplotstablenew[columns={x,y1,y2,sum}]{\pgfplotstablegetrowsof{\tableMean}}\tableStdp
			\pgfplotstablenew[columns={x,y1,y2,diff}]{\pgfplotstablegetrowsof{\tableMean}}\tableStdm
			\addplot[gray,very thick,mark=triangle] table{\tableMean};
			\addplot[name path=A,gray,dashed,forget plot] table [x=x, y=sum]{\tableStdp};
			\addplot[name path=B,gray,dashed,forget plot] table [x=x, y=diff]{\tableStdm};
			\addplot[gray!20,forget plot,opacity=0.5] fill between[of=A and B]; 
			\pgfplotstableread{images/esupp_AccAltProj_mean.dat}{\tableMean}
			\pgfplotstableread{images/esupp_AccAltProj_std.dat}{\tableStd}
   \pgfplotstableread{images/esupp_AccAltProj_std_low.dat}{\tableStdlow}
			\pgfplotstableset{
     			create on use/x/.style={create col/copy column from table={\tableMean}{x}}, 
     			create on use/y1/.style={create col/copy column from table={\tableMean}{y}},
     			create on use/y2/.style={create col/copy column from table={\tableStd}{y}}, 
        create on use/y3/.style={create col/copy column from table={\tableStdlow}{y}}, 
     			create on use/sum/.style={create col/expr={\thisrow{y1}+\thisrow{y2}}},
        create on use/diff/.style={create col/expr={\thisrow{y1}-\thisrow{y3}}}
			}
			\pgfplotstablenew[columns={x,y1,y2,sum}]{\pgfplotstablegetrowsof{\tableMean}}\tableStdp
			\pgfplotstablenew[columns={x,y1,y3,diff}]{\pgfplotstablegetrowsof{\tableMean}}\tableStdm
			\addplot[darkred,very thick,mark=square] table{\tableMean};
			\addplot[name path=A,darkred,dashed,forget plot] table [x=x, y=sum]{\tableStdp};
			\addplot[name path=B,darkred,dashed,forget plot] table [x=x, y=diff]{\tableStdm};
			\addplot[darkred!20,forget plot,opacity=0.5] fill between[of=A and B]; 
			\pgfplotstableread{images/esupp_Unrolled_mean.dat}{\tableMean}
			\pgfplotstableread{images/esupp_Unrolled_std.dat}{\tableStd}
            \pgfplotstableread{images/esupp_Unrolled_std_low.dat}{\tableStdlow}
			\pgfplotstableset{
     			create on use/x/.style={create col/copy column from table={\tableMean}{x}}, 
     			create on use/y1/.style={create col/copy column from table={\tableMean}{y}},
     			create on use/y2/.style={create col/copy column from table={\tableStd}{y}},
                    create on use/y3/.style={create col/copy column from table={\tableStdlow}{y}}, 
     			create on use/sum/.style={create col/expr={\thisrow{y1}+\thisrow{y2}}},     
     			create on use/diff/.style={create col/expr={\thisrow{y1}-\thisrow{y3}}}
			}
			\pgfplotstablenew[columns={x,y1,y2,sum}]{\pgfplotstablegetrowsof{\tableMean}}\tableStdp
			\pgfplotstablenew[columns={x,y1,y3,diff}]{\pgfplotstablegetrowsof{\tableMean}}\tableStdm
			\addplot[blue,very thick,mark=o] table{\tableMean};
			\addplot[name path=A,blue,dashed,forget plot] table [x=x, y=sum]{\tableStdp};
			\addplot[name path=B,blue,dashed,forget plot] table [x=x, y=diff]{\tableStdm};
			\addplot[blue!20,forget plot,opacity=0.5] fill between[of=A and B]; 
            
		\end{groupplot}
	\end{tikzpicture}
	\vspace{-0.5cm}
    \caption{Means and standard deviations  of  errors.  The ``Unrolled (st)'' curve stands for unrolled algorithm with {\em  soft-thresholding}. }
    \label{fig:compare_algs}
\end{figure*}
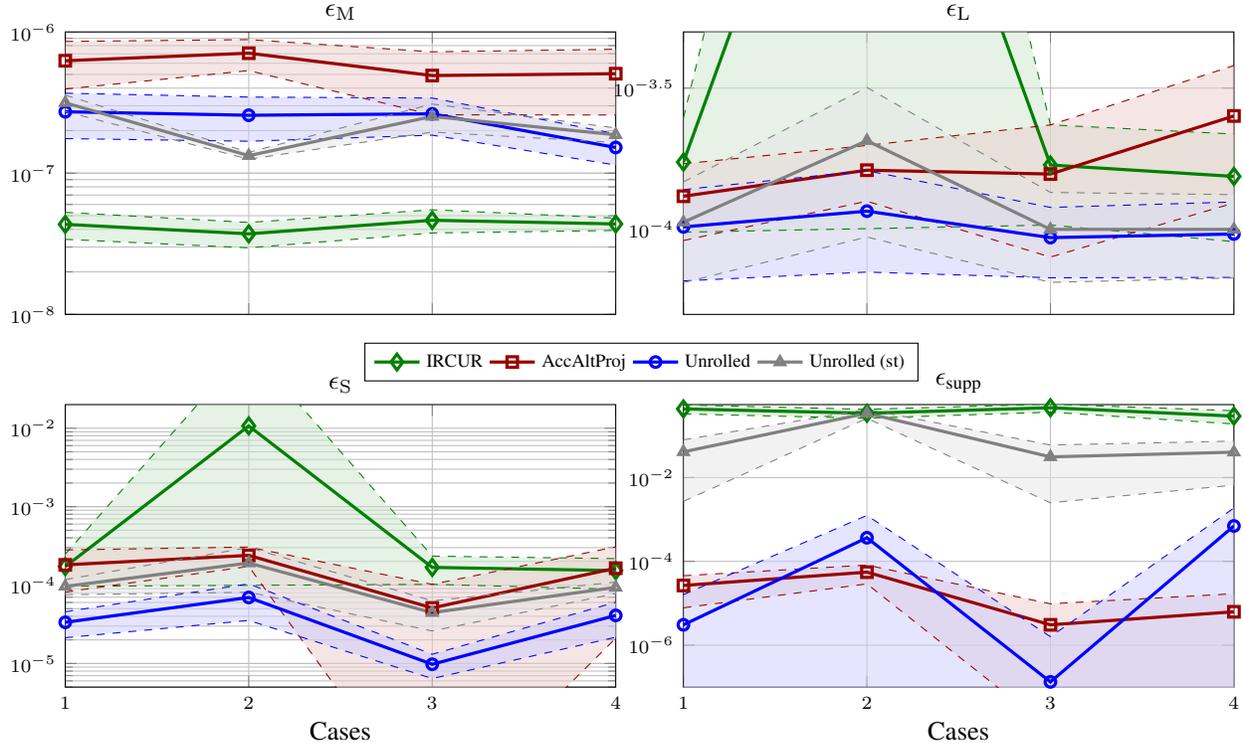

{\bf  Results:} In Fig.~\ref{fig:compare_algs}, we report  the four errors described in the four cases. We compare the performance of the proposed approach with IRCUR \cite{ircur:2021} and AccAltProj \cite{accaltproj:2019} (which are not unrolled algorithms).

We observe from Fig.~\ref{fig:compare_algs} that the proposed unrolled algorithm improves over its classical counterpart, which means that the   hyperparameters are learned well. The lowest error in  $\M$ is always achieved by the IRCUR method and remains small (order $10^{-8}$ versus $10^{-7}$) for the other methods. This can be explained by the fact that the hyperparameters are learnt so as to minimize the error on $\L$ and $\S$. This is confirmed by the results obtained individually on matrices $\L$ and $\S$ for which the smallest error is always obtained by the proposed unrolled method. Finally, as expected, the unrolled algorithm using the $\ell_1$-norm instead of the MCP does not perform well.

\begin{table}[!htb]
\small
\begin{center}
\renewcommand{\arraystretch}{1.1}
\begin{tabular}{|c|c|c|c|c|}
 \hline
   & Case 1 & Case 2 & Case 3 & Case 4 \\
 \hline
 $\!\gamma$ & $7.74 \!\times\! 10^{-1}$  & $7.71 \!\times\! 10^{-1}$  & $7.88 \!\times\! 10^{-1}$  & $7.40 \!\times\! 10^{-1}$  \\
\hline
 $\!\beta$ & $7.03  \!\times\! 10^{-2}$  &  $7.71 \!\times\! 10^{-2}$  & $5.00 \!\times\! 10^{-2}$  &  $4.58 \!\times\! 10^{-2}$  \\
\hline
\end{tabular}
\caption{Parameters learnt from the trained unrolled network in the different experimental settings. \label{tab:params}}
\end{center} \vspace{-.1in}
\end{table}

From Table \ref{tab:params}, we observe that the learned $\gamma$'s are similar across different settings, where they are all slightly larger than their initialised value  of $0.7$. This suggests that the default value of $\gamma = 0.7$ suggested in \cite{accaltproj:2019} is a fairly good estimate. The slight increase may be because  AccAltProj implements an early stopping criterion, where stops once the error $\frac{\|\mathbf{M}^\star - \mathbf{L}_k - \mathbf{S}_k\|_{\mathrm{F}}}{\|\mathbf{M}^\star\|_{\mathrm{F}}}$ at iteration $k < 50$ is below the  tolerance of $10^{-6}$. As such, for a larger fixed number of layers, a larger $\gamma$ would be needed so that the network converges more slowly to the same point. Conversely, the learned $\gamma$ would be smaller if we reduced the number of layers for the unrolled network. This means that learned $\gamma$ is optimised for the given fixed number of layers in the network.
The learned $\beta$ exhibits much more variation across the different cases. In particular, they increased by about 2 times from before training in Cases~1 and~2, and about 1.5 times in Cases 3 and 4. This observation is in line with the interpretation of $\beta$ in \cite{altproj:2014}, which stated that a higher value of $\beta$ results in $\hat{\mathbf{L}}$ that is more ``spiky" and $\hat{\mathbf{S}}$ that is more heavily diffused. We can take Case 1 as the baseline for the other cases to compare against. In Case 2 where $\alpha$ is greater than in Case 1, there would be more non-zero values in $\mathbf{S}^\star$, making it more diffused. In contrast, with smaller $\alpha$ in Case 3, the few non-zero elements of $\mathbf{S}^\star$ become more prominent against the backdrop of the other zero-valued elements, making it less diffused. In Case 4 where the magnitude of non-zero elements in $\mathbf{S}^\star$ is $10$ times of that in Case~1, the non-zero values are more pronounced and hence $\mathbf{S}^\star$ is less diffused. As the learned $\beta$ matches our expectation from theory for each case, this demonstrates that our unrolled model is indeed able to automatically fine-tune the parameter $\beta$ to the different settings, which is an advantage over the classical AccAltProj.

\vspace{-.07in}
\subsection{Face Dataset}
\vspace{-.07in}
{\bf Problem setup:} We now test the proposed unrolled model on the Yale Face Database \cite{yaledataset:1997} for the application of face modeling. The Yale database consists of 11 grayscale facial images each for a total of 15 subjects. The 11 images, each having a dimension of $243 \times 320$, show the same individual with different facial expressions, lighting conditions or accessories such as spectacles on the face. The task is of face modeling is to recover the occlusion-free image for facial recognition~\cite{facerec:2018}.

We vectorise the images of each subject, which are then stacked   to form a $77760 \times 11$ matrix $\M^\star$. The static occlusion-free image of the subject forms the low-rank component of the matrix while the varied facial expressions, shadows, and objects covering the face form the low-rank component. Since all 11 images  share one common underlying occlusion-free facial image, we   assume that $\mathrm{rank}(\M^\star) = 1$.

Subjects 1 to 7  and Subjects 8 to 15 are used for training and testing respectively. Similar to the  experiment on synthetic datasets, we use IRCUR to obtain initial estimates  and train the unrolled network for 8 epochs.

\begin{figure}[!ht]
\centering
\begin{tabular}{p{0cm}cc}
& Subject 10 & Subject 13\\
\rotatebox[origin=l]{90}{\qquad \quad IRCUR}  & \includegraphics[trim={2cm 0 2cm 1cm},clip, width=3.7cm]{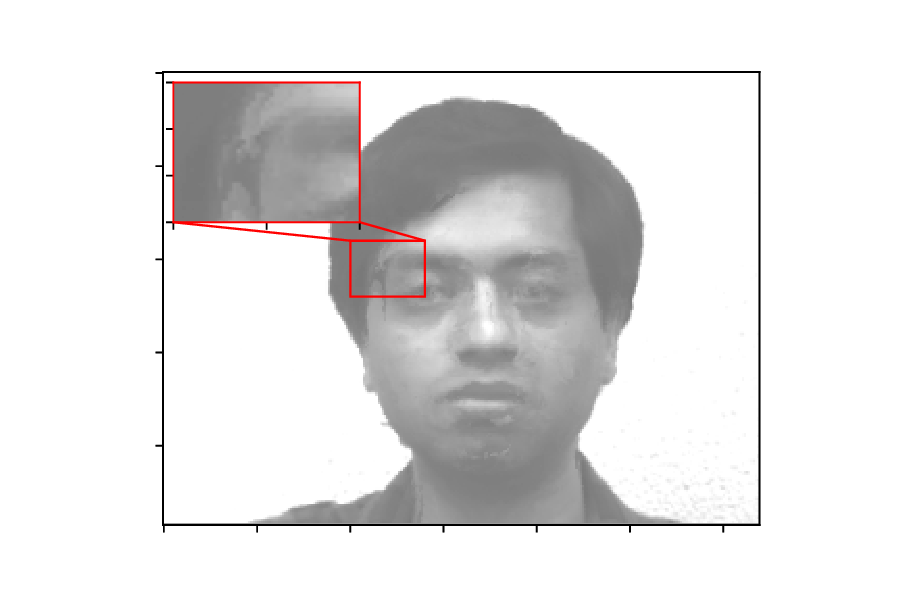}
&
\includegraphics[trim={2cm 0 2cm 1cm},clip, width=3.7cm]{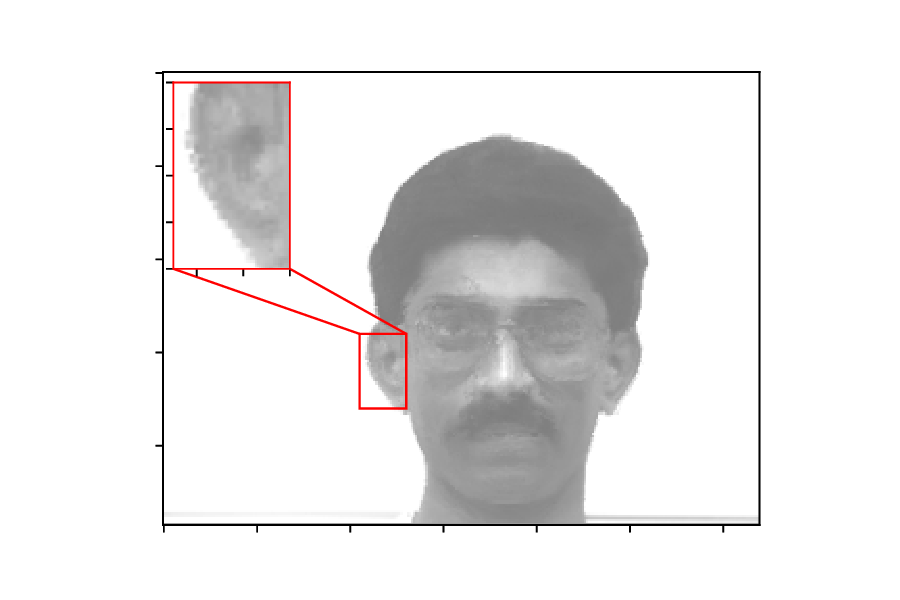}\\
\rotatebox[origin=l]{90}{\qquad  AccAltProj} & \includegraphics[trim={2cm 0 2cm 1cm},clip, width=3.7cm]{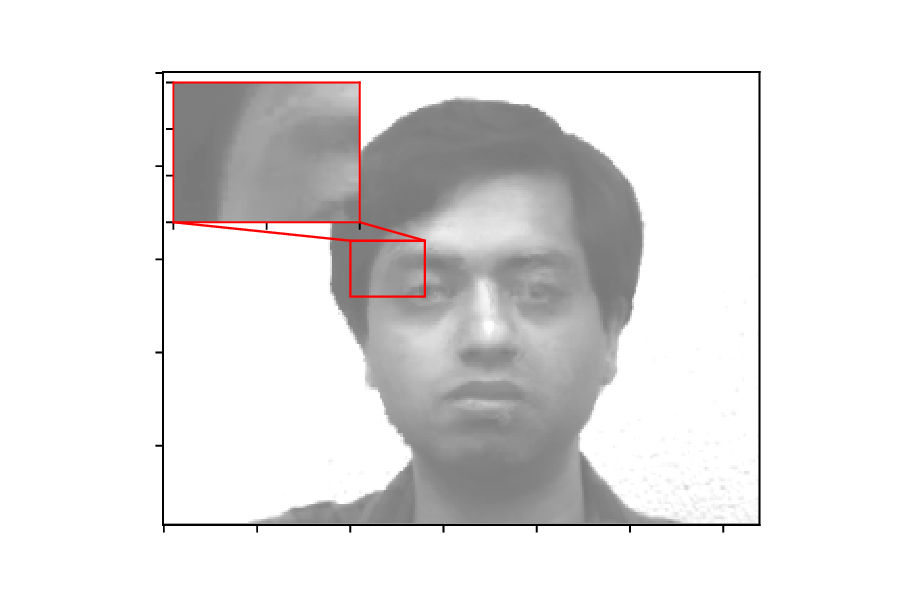}
&
\includegraphics[trim={2cm 0 2cm 1cm},clip, width=3.7cm]{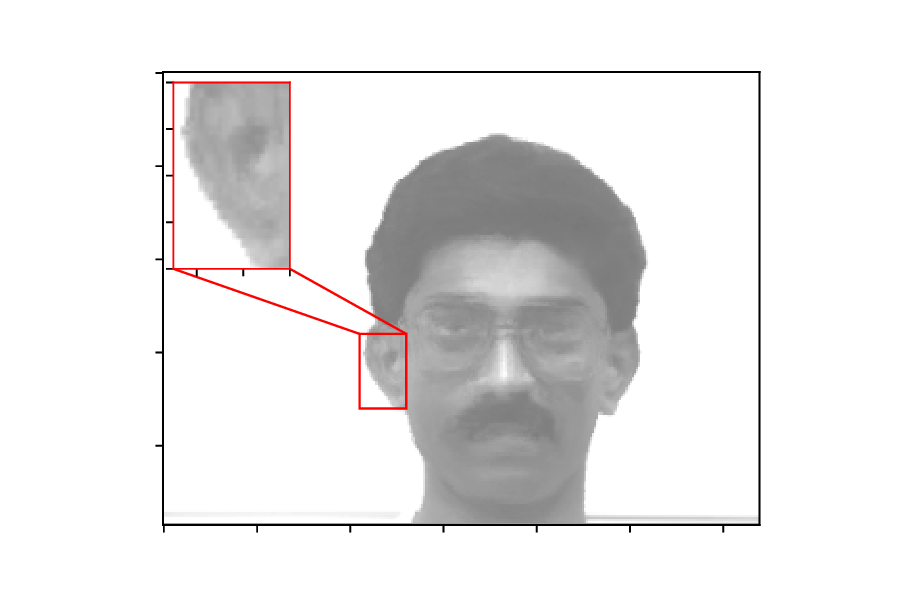}\\
\rotatebox[origin=l]{90}{\qquad \quad Unrolled} &
\includegraphics[trim={2cm 0 2cm 1cm},clip, width=3.7cm]{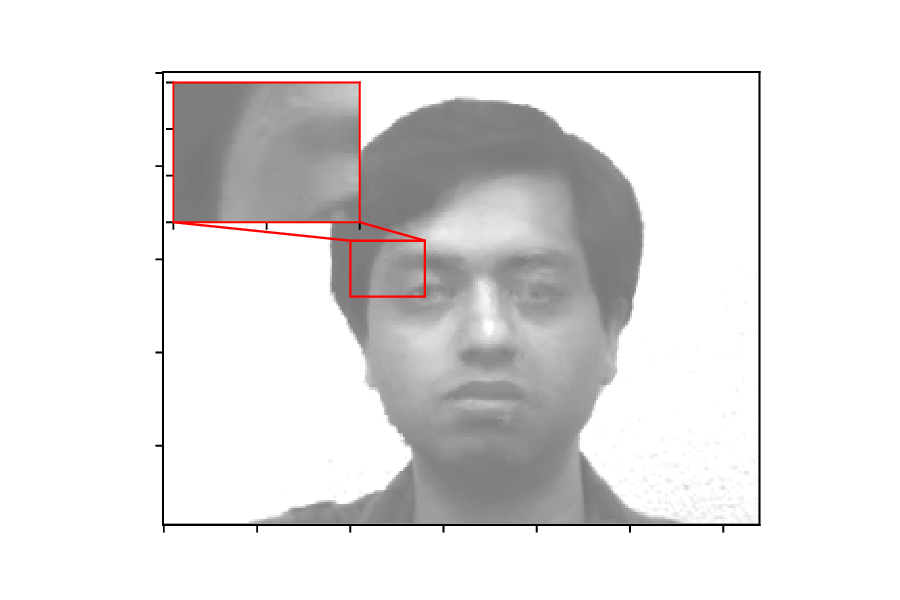}
&
\includegraphics[trim={2cm 0 2cm 1cm},clip, width=3.7cm]{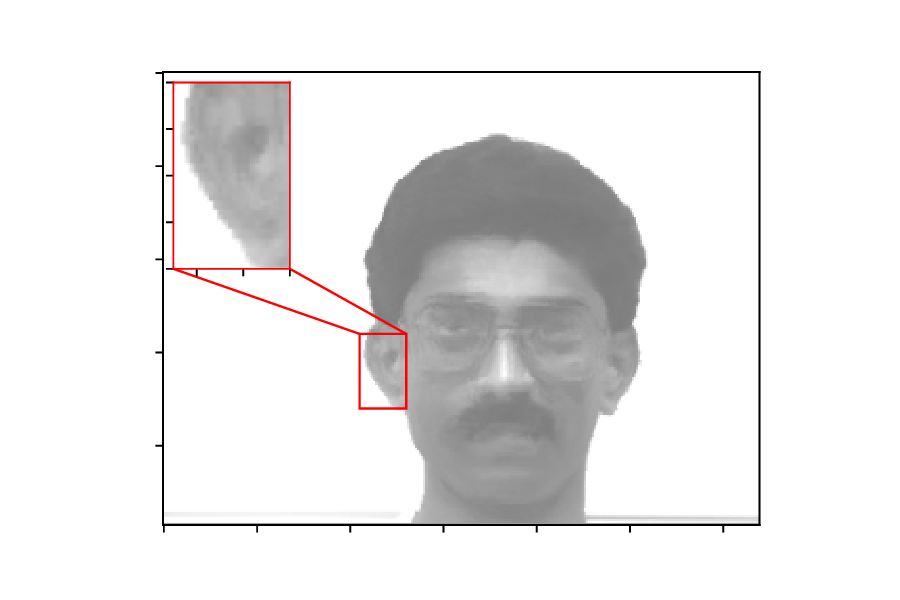}\\
\end{tabular}
\caption{Low-rank components of subjects 10 and 13 recovered from IRCUR, AccAltProj, and   unrolled network. AccAltProj and its unrolled version result in images with sharper edges.}
\label{fig:res_lr}
\end{figure}

\begin{figure}[!ht]
\centering
\begin{tabular}{p{0cm}cc}
& Subject 10 & Subject 13\\
\rotatebox[origin=l]{90}{\qquad \quad IRCUR}  & \includegraphics[trim={2cm 0 2cm 1cm},clip, width=3.7cm]{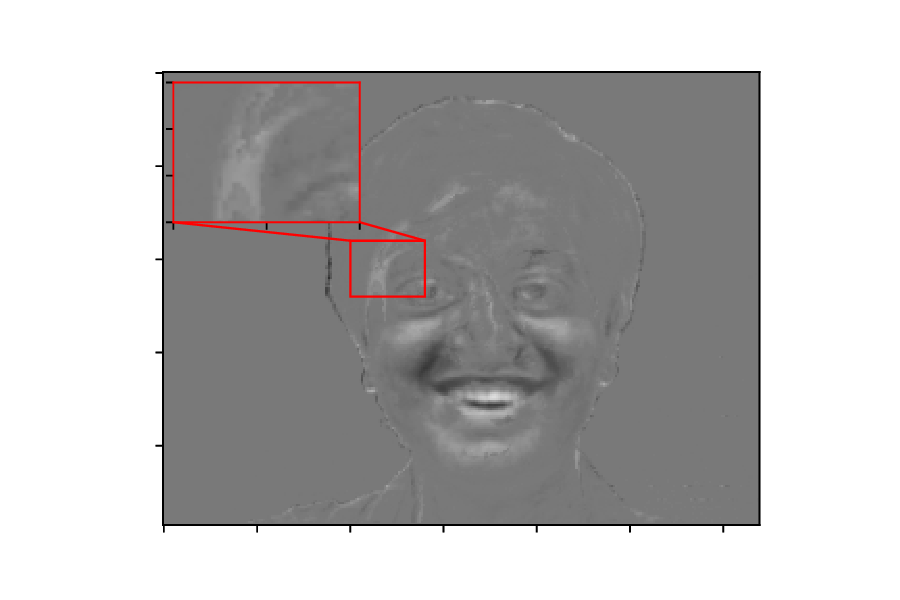}
&
\includegraphics[trim={2cm 0 2cm 1cm},clip, width=3.7cm]{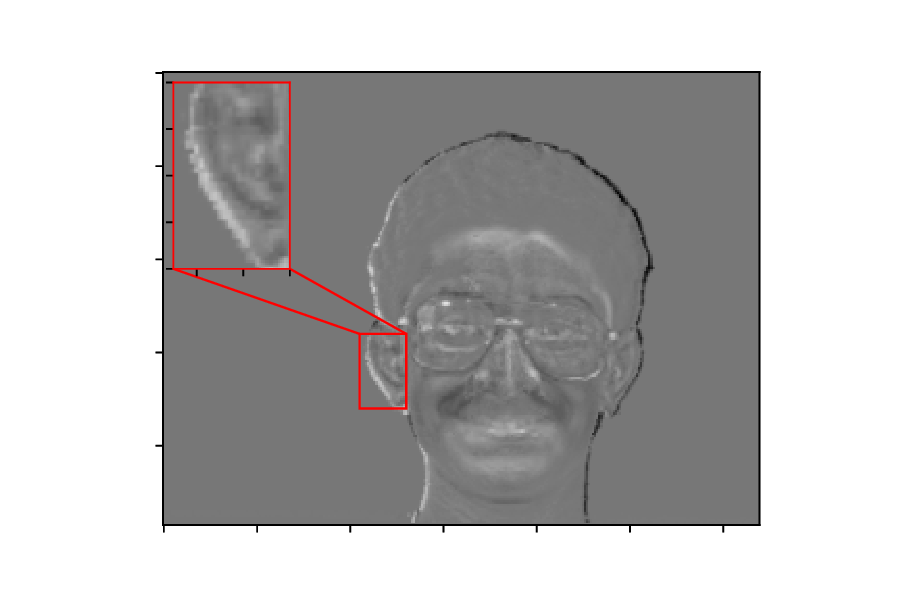}\\
\rotatebox[origin=l]{90}{\qquad  AccAlterProj} & \includegraphics[trim={2cm 0 2cm 1cm},clip, width=3.7cm]{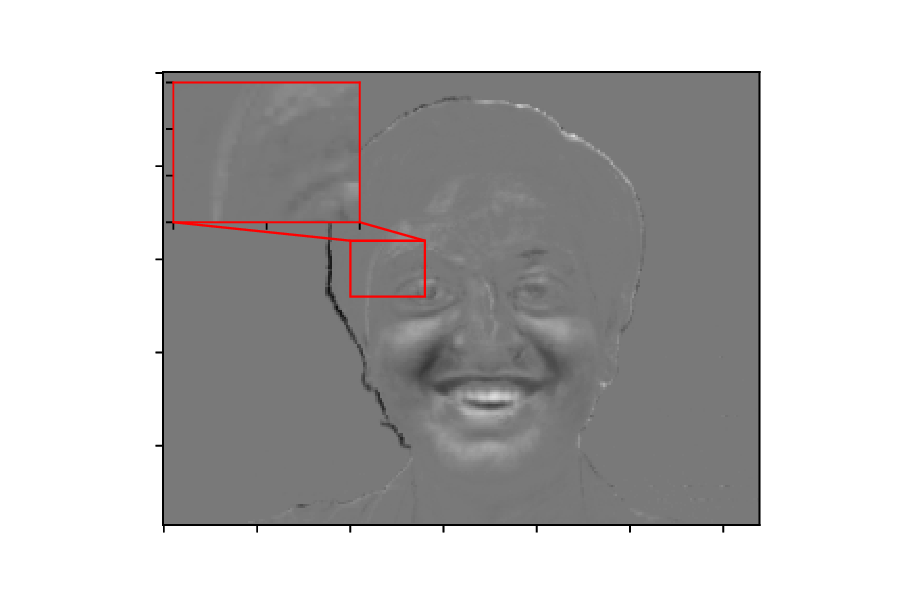}
&
\includegraphics[trim={2cm 0 2cm 1cm},clip, width=3.7cm]{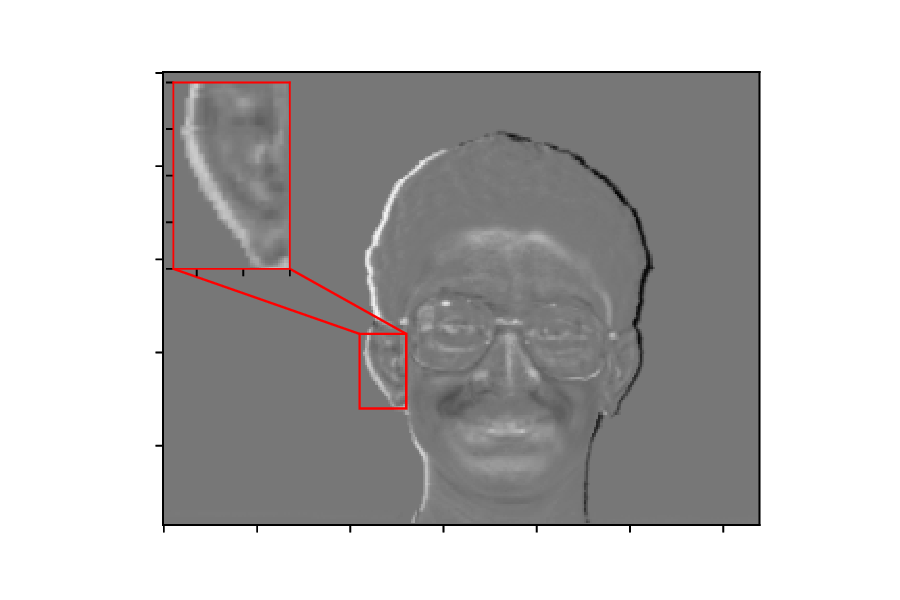}\\
\rotatebox[origin=l]{90}{\qquad \quad Unrolled} &
\includegraphics[trim={2cm 0 2cm 1cm},clip, width=3.7cm]{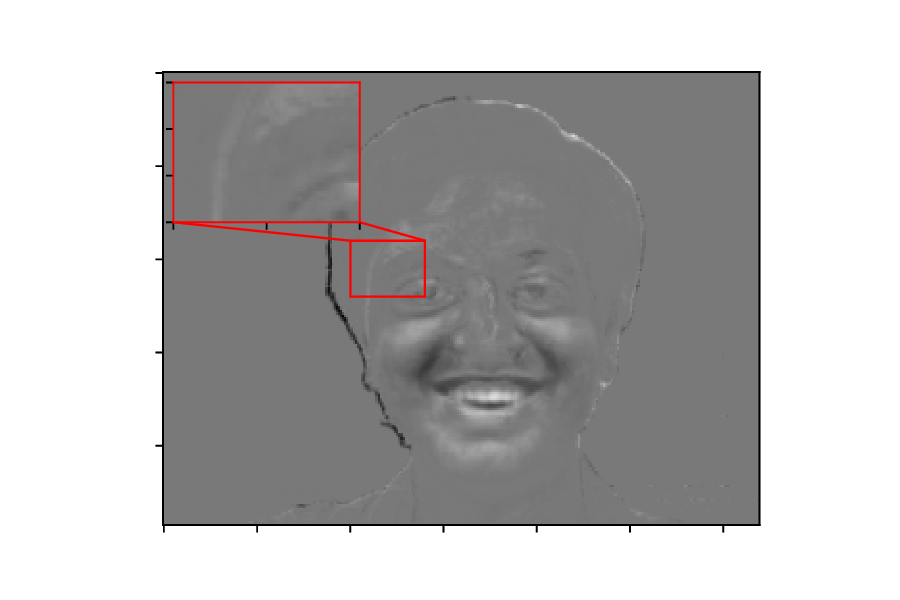}
&
\includegraphics[trim={2cm 0 2cm 1cm},clip, width=3.7cm]{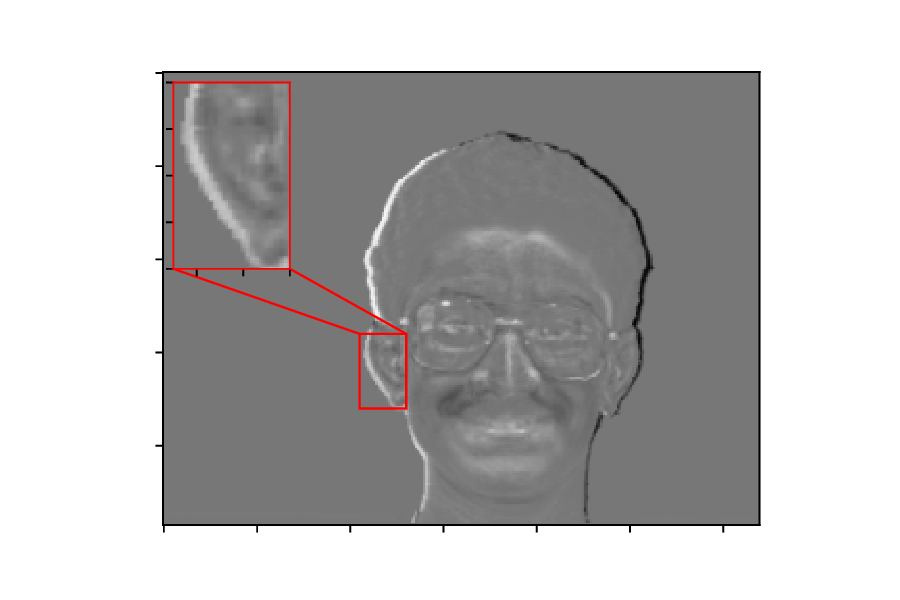}\\
\end{tabular}
\caption{Sparse components (happy) of subjects 10 and 13 recovered from IRCUR, AccAltProj and   unrolled network. Observe that AccAltProj and its unrolled version result in smoother estimates while preserving details.}
\label{fig:res_sp}
\end{figure}

{\bf Results:}
Visual results are displayed in Figs.~\ref{fig:res_lr} and~\ref{fig:res_sp} for the low rank and  sparse parts respectively. The methods enable the separation of the original images into expressionless faces and expression details. While the images learned by  IRCUR  are poor, the proposed unrolled strategy adaptively learned  hyperparameters $(\gamma,\beta)$ that result in sharper edges. 
 Tests were performed on an Intel(R) Core(TM) i7-1185G7 @3.00GHz, with 32Go RAM. The total (7 subjects) training time is 300s. Testing times (per subject on average) are as follows: IRCUR: 8s, AccAltProj: 0.75s and unrolled procedure:
4.375s, showing that the unrolled procedure has a good accuracy-computation time
tradeoff.



\vspace{-.1in}

\section{Conclusion}
\label{sec:conclu}\vspace{-.1in}
We proposed an unrolled algorithm to solve the RPCA problem in its nonconvex form. This results in an unrolled version of the AccAltProj algorithm but incorporates the Minimax Concave Penalty. The underlying learning strategy, which has the advantage of learning hyperparameters $\gamma$ and $\beta$ automatically, allows us to improve the state-of-the-art performances on benchmark synthetic datasets used in existing works as well as on real-world  face datasets. In future work, we plan to improve on the training criterion in Section~\ref{ssec:train_crit}  as well as the automatic learning of more parameters such as the ones that parametrize the MCP, i.e., $\zeta$ and $\upsilon$ in~\eqref{eq:MCP}.


\bibliographystyle{IEEEbib}
\bibliography{abbr,references}

\begin{thebibliography}{10}

\bibitem{candes2011robust}
E.~J. Cand{\`e}s, X.~Li, Y.~Ma, and J.~Wright,
\newblock ``Robust principal component analysis?,''
\newblock {\em J. ACM}, vol. 58, no. 3, pp. 1--37, 2011.

\bibitem{deerwester1990indexing}
S.~Deerwester, S.~T. Dumais, G.~W. Furnas, T.~K. Landauer, and R.~Harshman,
\newblock ``Indexing by latent semantic analysis,''
\newblock {\em J. Am. Soc. Inf. Sci.}, vol. 41, no. 6, pp. 391--407, 1990.

\bibitem{facerec:2018}
T.~Bouwmans, S.~Javed, H.~Zhang, Z.~Lin, and R.~Otazo,
\newblock ``On the applications of robust {PCA} in image and video
  processing,''
\newblock {\em Proc. IEEE}, vol. 106, no. 8, pp. 1427--1457, 2018.

\bibitem{chandrasekaran}
V.~Chandrasekaran, P.~A. Parrilo, and A.~S. Willsky,
\newblock ``Latent variable graphical modeling via convex optimization,''
\newblock {\em The Annals of Statistics}, vol. 40, no. 4, pp. 1935--1967, 2012.

\bibitem{koren2021advances}
Y.~Koren, S.~Rendle, and R.~Bell,
\newblock ``Advances in collaborative filtering,''
\newblock {\em Recommender systems handbook}, pp. 91--142, 2021.

\bibitem{rpcaconvex:2009}
J.~Wright, A.~Ganesh, S.~Rao, Y.~Peng, and Y.~Ma,
\newblock ``Robust principal component analysis: Exact recovery of corrupted
  low-rank matrices via convex optimization,''
\newblock in {\em Adv. Neural Inf. Process. Syst.}, Y.~Bengio, D.~Schuurmans,
  J.~Lafferty, C.~Williams, and A.~Culotta, Eds. 2009, vol.~22, Curran
  Associates, Inc.

\bibitem{rpcaconvexlista:2009}
Z.~Lin, A.~Ganesh, J.~Wright, L.~Wu, M.~Chen, and Y.~Ma,
\newblock ``Fast convex optimization algorithms for exact recovery of a
  corrupted low-rank matrix,''
\newblock {\em Proc. IEEE Int. Workshop on Computational Advances in
  Multi-Sensor Adaptive Processing (CAMSAP)}, vol. 61, 2009.

\bibitem{altproj:2014}
P.~Netrapalli, U.~N. Niranjan, S.~Sanghavi, A.~Anandkumar, and P.~Jain,
\newblock ``Non-convex robust {PCA},'' 2014.

\bibitem{accaltproj:2019}
H.~Cai, J.-F. Cai, and K.~Wei,
\newblock ``Accelerated alternating projections for robust principal component
  analysis,''
\newblock {\em J. Mach. Learn. Res.}, vol. 20, no. 1, pp. 685–717, Jan 2019.

\bibitem{ircur:2021}
H.~Cai, K.~Hamm, L.~Huang, J.~Li, and T.~Wang,
\newblock ``Rapid robust principal component analysis: {CUR} accelerated
  inexact low rank estimation,''
\newblock {\em IEEE Signal Process. Lett.}, vol. 28, pp. 116--120, Feb 2021.

\bibitem{Monga2021}
V.~Monga, Y.~Li, and Y.~Eldar,
\newblock ``Algorithm unrolling: Interpretable, efficient deep learning for
  signal and image processing,''
\newblock {\em IEEE Signal Process. Mag.}, vol. 38, no. 2, pp. 18--44, 3 2021.

\bibitem{corona:2020}
O.~Solomon, R.~Cohen, Y.~Zhang, Y.~Yang, Q.~He, J.~Luo, R.~J.~G. van Sloun, and
  Y.~C. Eldar,
\newblock ``Deep unfolded robust {PCA} with application to clutter suppression
  in ultrasound,''
\newblock {\em IEEE Trans. Med. Imag.}, vol. 39, no. 4, pp. 1051--1063, 2020.

\bibitem{refrpca:2021}
H.~Van Luong, B.~Joukovsky, Y.~C. Eldar, and N.~Deligiannis,
\newblock ``A deep-unfolded reference-based {RPCA} network for video
  foreground-background separation,''
\newblock {\em Proc. Eur. Sig. Image Proc. Conf.}, pp. 1432--1436, 2021.

\bibitem{solomon:2019}
R.~Cohen, Y.~Zhang, O.~Solomon, D.~Toberman, L.~Taieb, R.~J.~G. van Sloun, and
  Y.~C. Eldar,
\newblock ``Deep convolutional robust {PCA} with application to ultrasound
  imaging,''
\newblock in {\em Proc. Int. Conf. Acoust. Speech Signal Process.}, 2019, pp.
  3212--3216.

\bibitem{ultrasound:2020}
R.~J.~G. van Sloun, R.~Cohen, and Y.~C. Eldar,
\newblock ``Deep learning in ultrasound imaging,''
\newblock {\em Proc. IEEE}, vol. 108, no. 1, pp. 11--29, 2020.

\bibitem{xray:2022}
B.~Qin, H.~Mao, Y.~Liu, J.~Zhao, Y.~Lv, Y.~Zhu, S.~Ding, and X.~Chen,
\newblock ``Robust {PCA} unrolling network for super-resolution vessel
  extraction in x-ray coronary angiography,''
\newblock {\em IEEE Trans. Med. Imag.}, vol. 41, no. 11, pp. 3087--3098, 2022.

\bibitem{multimodalrpca:2022}
S.~Markowitz, C.~Snyder, {Y. C.} Eldar, and {M. N.} Do,
\newblock ``Multimodal unrolled robust {PCA} for background foreground
  separation,''
\newblock {\em IEEE Trans. Image Process.}, vol. 31, pp. 3553--3564, 2022.

\bibitem{lrpca:2021}
H.~Cai, J.~Liu, and W.~Yin,
\newblock ``Learned robust {PCA}: A scalable deep unfolding approach for
  high-dimensional outlier detection,''
\newblock in {\em Adv. Neural Inf. Process. Syst.}, A.~Beygelzimer, Y.~Dauphin,
  P.~Liang, and J.~Wortman Vaughan, Eds., 2021, pp. 16977--16989.

\bibitem{lista:2010}
K.~Gregor and Y.~LeCun,
\newblock ``Learning fast approximations of sparse coding,''
\newblock in {\em Proc. Int. Conf. Mach. Learn.}, Madison, WI, USA, 2010,
  ICML'10, p. 399–406, Omnipress.

\bibitem{mcp:2010}
C.-H. Zhang,
\newblock ``{Nearly unbiased variable selection under minimax concave
  penalty},''
\newblock {\em Ann. Statist.}, vol. 38, no. 2, pp. 894 -- 942, 2010.

\bibitem{unifiedl2l0}
E.~Soubies, L.~Blanc-F{\'e}raud, and G.~Aubert,
\newblock ``A unified view of exact continuous penalties for l2-l0
  minimization,''
\newblock {\em SIAM J. Optim.}, vol. 27, no. 3, pp. 2034--2060, 2017.

\bibitem{firmthres:2016}
J.~Woodworth and R.~Chartrand,
\newblock ``Compressed sensing recovery via nonconvex shrinkage penalties,''
\newblock {\em Inverse Problems}, vol. 32, no. 7, pp. 075004, may 2016.

\bibitem{yaledataset:1997}
P.~N. Belhumeur, J.~P. Hespanha, and D.~J. Kriegman,
\newblock ``Eigenfaces vs. fisherfaces: recognition using class specific linear
  projection,''
\newblock {\em IEEE Trans. Pattern Anal. Mach. Intell.}, vol. 19, no. 7, pp.
  711--720, 1997.

\end{thebibliography}

\end{document}